\documentclass[preprint,prb,aps,altaffilletter]{revtex4}

\usepackage{graphicx}

\usepackage{dcolumn}
\usepackage{bm}

\begin{document}

\title{A widely tunable parametric amplifier based on a SQUID array resonator}

\author{M.~A.~Castellanos-Beltran}
\email[Electronic mail: ]{castellm@colorado.edu}
\author{K.~W.~Lehnert}
\affiliation{JILA, National Institute of Standards and Technology
and the University of Colorado, and Department of Physics,
University of Colorado, Boulder, Colorado 80309}
\date{\today}

\begin{abstract}

We create a Josephson parametric amplifier from a transmission line
resonator whose inner conductor is made from a series SQUID array.
By changing the magnetic flux through the SQUID loops, we are able
to adjust the circuit's resonance frequency and, consenquently, the
center of the amplified band, between 4 and 7.8~GHz. We observe that
the amplifier has gains as large as 28~dB and infer that it adds
less than twice the input vacuum noise.
\\

\end{abstract}

\maketitle

Josephson parametric amplifiers (JPAs) operate as ultralow noise
microwave amplifiers. By detecting only one quadrature of a signal,
degenerate parametric amplifiers can add even less noise than the
minimum required by quantum mechanics when detecting both
quadratures.\cite{takahasi1965} Josephson parametric amplifiers have
been operated with near quantum-limited sensitivity\cite{yurke1989}
and have been used to squeeze both thermal and vacuum
noise.\cite{yurke1987,yurke1988,yurke1990} In spite of these
promising results, they have not been widely adopted since they
suffer from two main disadvantages. They have both small dynamic
range and narrow band gain; they are well suited to amplify signals
only in a narrow range in both power and frequency, limiting their
application. Furthermore, compelling applications that would benefit
from a lower-noise microwave amplifier have only recently been
developed. With the advent of quantum information processing using
superconducting circuits,\cite{wallraff2004,schuster2007} there is
now a need for practical amplifiers that operate at the limits
imposed by quantum mechanics.

The crucial element in a resonant-mode parametric amplifier is a
circuit whose resonance frequency can be varied with time. If a
reactive parameter oscillates at twice the resonance frequency,
energy can be pumped into (or out of) the mode, realizing an
amplifier. In practice, this time dependence is often generated
through a nonlinear inductance or capacitance. If the nonlinear
reactance is proportional to the intensity rather than the amplitude
of the mode, then an intense pump tone applied at the resonance
frequency $\omega$ automatically creates the necessary $2\omega$
parametric oscillation. In analogy with optics, we describe this
effect as a Kerr nonlinearity. The nonlinear current-dependent
inductance of a Josephson junction,
\begin{equation}
  L_{j}(I) =\frac{\hbar}{2eI_c}\frac{\arcsin{(I/I_{c}})}{I/I_{c}}
  \label{eq:L}
\end{equation}
provides such a Kerr nonlinearity, where $I_{c}$ is the critical
current of the junction, and $I$ is the current flowing through it.

Because they are built from nonlinear resonant circuits, Josephson
parametric amplifiers are inherently narrowband with limited dynamic
range. Only signals close to the circuit's resonance frequency whose
power is small compared to the pump can be linearly amplified. In
this paper, we report a novel approach that addresses the limited
bandwidth of Josephson parametric amplifiers. We create a JPA from a
circuit whose resonance frequency can be adjusted between 4 and 7.8
GHz by applying a magnetic field. The amplifier is still narrowband,
but the band center can be adjusted over an octave in frequency.
With the amplifier, we demonstrate power gains as large as $28$ dB.
Furthermore, we can extract the amplifier parameters by measuring
the reflectance from the resonator and use them to accurately
predict the amplifier's frequency-dependent gain. Finally, the
sensitivity is improved by 16~dB when we place our parametric
amplifier in front of a state-of-the-art microwave amplifier (HEMT).
This improvement demonstrates that the parametric amplifier provides
useful gain and operates much closer to the quantum limit than the
HEMT amplifier.

\begin{figure}
\includegraphics{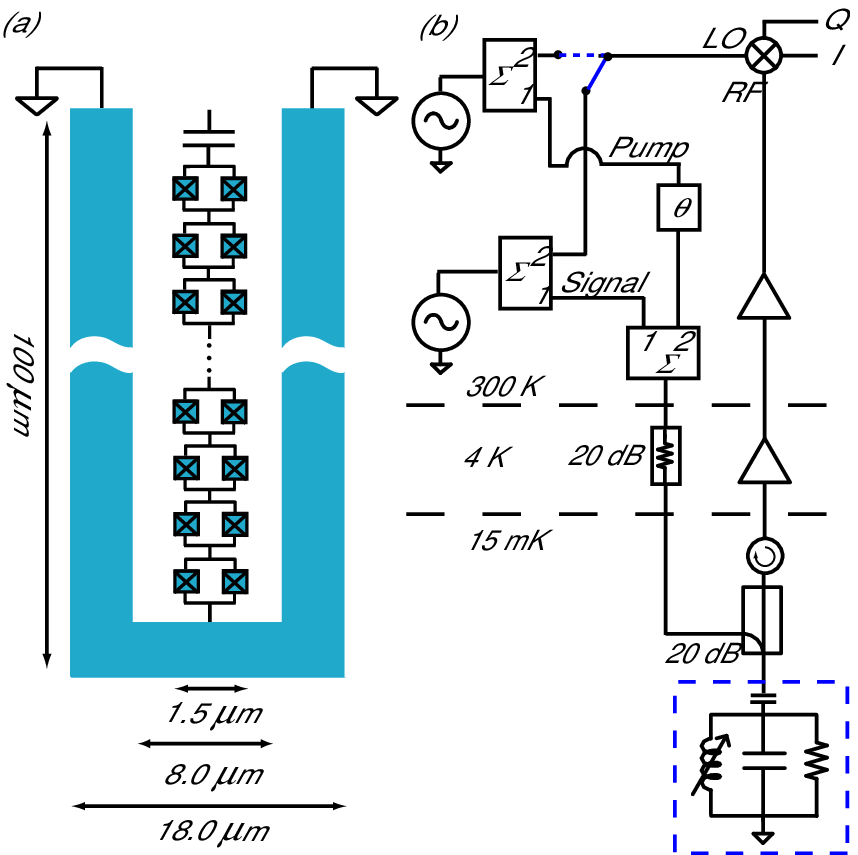}
\caption{\label{fig:Fig1} Device diagram and measurement schematic.
(a) The device's center conductor is a series array of 400 SQUIDs.
The resonator's ground plane is made out of aluminum, and the SQUIDs
are made from Al/AlOx/Al junctions. They are fabricated using E-beam
lithography and double angle evaporation on an oxidized silicon
substrate. (b) Simplified measurement schematic. We model the
resonator as an RLC circuit, as shown in the dashed box.}
\end{figure}

The device we study consists of a quarter-wave coplanar-waveguide
(CPW) resonator whose center conductor is an array of SQUIDs in
series [Fig. \ref{fig:Fig1}(a)]. Two Josephson junctions in parallel
form a SQUID, which behaves as a single junction with an effective
$I_c=I_{c}^{0}|\cos{(\pi\Phi/\Phi_0)}|$, where $\Phi/\Phi_0$ is the
magnetic flux enclosed by the SQUID loop in units of flux quanta. By
adjusting $\Phi$ through the SQUIDs, we can adjust the inductance
per unit length of the coplanar waveguide (Eq.~\ref{eq:L}).
\cite{haviland1996} We estimate $I_c^0$ for one SQUID to be $1.5\
\mu$A. The resulting metamaterial has a zero-flux inductance per
unit length of $L_l=0.9$~mH/m~$\approx 700\mu_0 $. The CPW has a
capacitance per unit length of $C_l=0.11$ nF/m, yielding a phase
velocity of $v_{ph}=1/\sqrt{L_lC_l}=0.01c$. We form a $\lambda/4$
resonator by shorting one end of the SQUID array CPW and
capacitively coupling the other end to a $50\ \Omega$ transmission
line. The SQUID array behaves as a lumped element
resonator\cite{feldman1975,wahlsten1977} close to its resonance
frequency; it is not a distributed parametric
amplifier.\cite{sweeny1985,yurke1996}

The parametric amplifier is operated in reflection mode, as shown in
Fig. \ref{fig:Fig1}(b). Two signal generators create two tones, a
pump at frequency $f_p$ and a signal at $f_s$. The two tones are
summed before being injected into a dilution refrigerator operating
at $15$~mK. They are attenuated by $20$~dB at $4$~K. A directional
coupler at $15$~mK provides an additional $20$~dB of attenuation and
separates the incident tones from the reflected tones. Thus,
including the 8-12~dB of loss from cables, incident tones and
room-temperature Johnson noise are attenuated by about $50$~dB.

Because of the nonlinearity of the metamaterial, the pump and signal
tones mix. This mixing amplifies the signal and creates an idler, or
intermodulation tone, at a frequency $f_{I}=2f_p-f_s$. To further
amplify the signals coming out of our resonator, we use a cryogenic
HEMT amplifier with noise temperature $T_{N}=5$~K and another set of
low noise amplifiers at room temperature. An isolator at base
temperature prevents the noise emitted by the input of the HEMT
amplifier from exciting the JPA. Amplitudes and phases of the
signals at the output of the room temperature amplifiers are
recovered with an IQ demodulator whose local oscillator (LO) can be
provided by either microwave generator.

\begin{figure}
\includegraphics{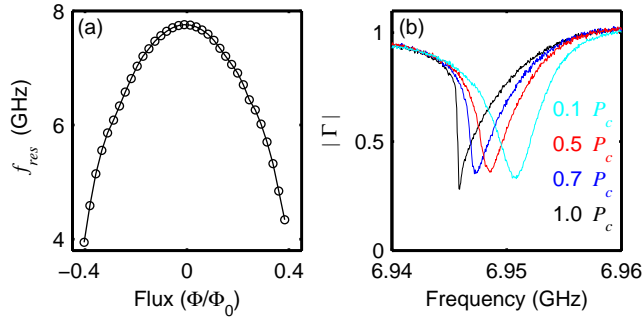}
\caption{\label{fig:Fig2} Flux and power dependance of the resonance
circuit. (a) Resonance frequency as a function of flux. (b)
Reflection coefficient magnitude as a function of frequency at
different pump powers for $\Phi=0.2\Phi_0$.}
\end{figure}

Before operating the parametric amplifier, we characterize the
resonator's reflectance with just a pump tone. We first study the
flux dependence of the resonator by measuring the real (I) and
imaginary (Q) part of the reflection coefficient $\Gamma$ as a
function of frequency. The resonance frequency $f_{res}$ is
identified by a dip in $\left|\Gamma\right|$. Figure
\ref{fig:Fig2}(a) shows how the resonance frequency behaves as a
function of $\Phi/\Phi_0$. The applied flux increases $L_l$,
reducing $v_{ph}$ and consequently $f_{res}$.

By measuring $\Gamma$ as a function of frequency and incident power,
we obtain the linear and nonlinear resonator parameters [Fig.
\ref{fig:Fig2}(b)]. At low enough incident power $P$, where the
resonator response is linear, we extract the damping rates
associated with the coupling capacitor $\gamma_{1}$ and the linear
dissipation in the resonator $\gamma_{2}$. We extract these from the
halfwidth [$(\gamma_1+\gamma_2)/2\pi$] and the depth of the dip
[$(\gamma_2-\gamma_1)/(\gamma_2+\gamma_1)$] in $\left|\Gamma\right|$
at the resonance frequency $\omega_0=2\pi f_{res}$. For a flux of
$\Phi=0.2\Phi_0$, we find the resonator's linear parameters
$f_{res}=6.952$~GHz, $\gamma_1/2\pi=1.9$ MHz, and
$\gamma_2/2\pi=1.1$ MHz. As we increase the pump power, the Kerr
nonlinearity makes the resonance frequency decrease according to the
equation $\omega_0-\omega_m+KE=0$, where $\omega_m$ is the frequency
at which $\Gamma$ is minimum, $K$ is the Kerr constant, and $E$ is
the energy stored in the resonator.\cite{yurke2006} Above the
critical power $P_c$, $\Gamma$ is discontinuous, and the resonators
response is bistable. From the frequency and power dependence of
$\Gamma$, we estimate the critical power and Kerr constant to be
$P_c=3\pm 1.3$~fW and $\hbar K=-1.6 \pm0.7 \times10^{-5}$,
respectively. The large uncertainty comes from the $4$~dB
uncertainty of the incident power on the resonator. From Appendix A
in Ref. \onlinecite{yurke2006}, we can calculate the Kerr constant
from the number of SQUIDs and their $I_c$. The expected value for
the Kerr constant is $\hbar K=-1\times10^{-5}$, in agreement with
our measurement. To model more completely the behavior of the
resonator, we also include a nonlinear dissipation term
$\gamma_3=0.027\left|K\right|$, which is the imaginary part of the
Kerr constant. From the physical characteristics of the resonator,
we can predict $f_{res}$, $\gamma_1$, and $K$; however we do not yet
understand the physical origin of $\gamma_2$ and $\gamma_3$.

The analysis of the parametric amplifier follows closely the theory
developed by Yurke and Buks for parametric amplification in
superconducting resonators.\cite{yurke2006} In their model, the Kerr
nonlinearity is provided by the intrinsic kinetic inductance of a
superconducting film,\cite{haviland2007} while in our case it arises
from the nonlinear Josephson inductance of the SQUIDs (Eq.
\ref{eq:L}).

\begin{figure}
\includegraphics{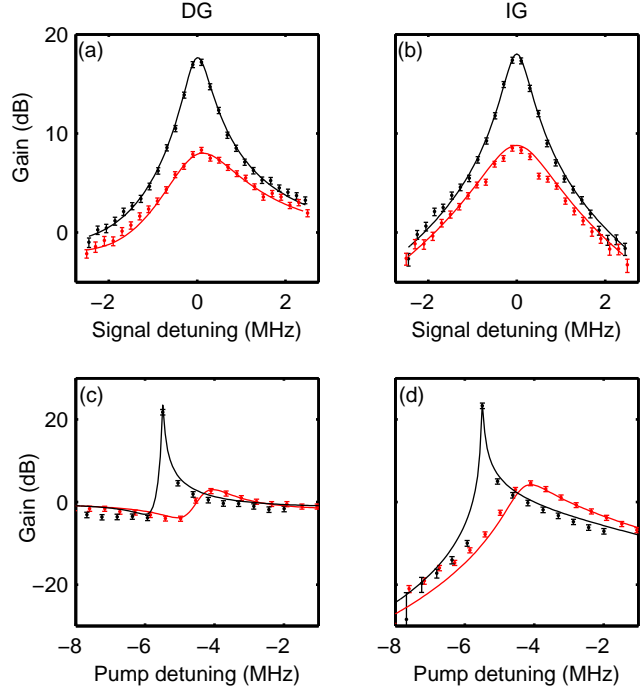}
\caption{\label{fig:Fig3} Performance of the amplifier in
nondegenerate mode. (a) and (b) DG and IG as functions of signal
detuning (points) and predictions of Ref. \onlinecite{yurke2006}
(lines) for two different pump powers. DG and IG for $P=0.9\ P_c$
(black) and for $P=0.75\ P_c$ (red). (c) and (d) DG and IG as
functions of pump detuning (points) for $P=0.95\ P_c$ (blue) and
$P=0.5\ P_c$ (red) and prediction (lines) of Ref.
\onlinecite{yurke2006}.}
\end{figure}

The intermodulation gain (IG) and direct gain (DG) can be predicted
from the resonator's parameters. We define DG as the ratio between
the reflected signal power with the pump on and the incident signal
power; IG is the ratio between the intermodulation tone and the
incident signal. To verify the behavior of the parametric amplifier,
we operate it in the nondegenerate mode and measure the frequency
dependence of both gains in two different ways. In the nondegenerate
mode, the signal and the pump frequencies are different, and the
generator that creates the signal tone also provides the LO to the
demodulator. In the first test, we apply the pump at a frequency
close to $\omega_m$ and analyze DG and IG as we detune the signal
frequency from the pump by an amount $\delta$$f$. In Figs.
\ref{fig:Fig3}(a) and \ref{fig:Fig3}(b), we plot both IG and DG as a
function $\delta$$f$ for two different pump powers. We also plot the
predictions from the theory in Ref. \cite{yurke2006} where the
parameters in the theory are extracted from the measurements of
$\Gamma$. From this plot, we estimate the $3$ dB bandwidth to be
about $300$ kHz when DG and IG are $18$~dB. Next, we measure IG and
DG as a function of pump detuning, \textit{i.e.}, the difference in
frequency between the applied pump and $f_{res}$. In this test, the
signal and the pump frequency differ by a fixed amount,
$\delta$$f=10$~kHz, [Figs. \ref{fig:Fig3}(c) and \ref{fig:Fig3}(d)].
From the agreement seen in Fig. \ref{fig:Fig3}, we conclude that
Ref. \onlinecite{yurke2006} provides an appropriate model for our
device.

\begin{figure}
\includegraphics{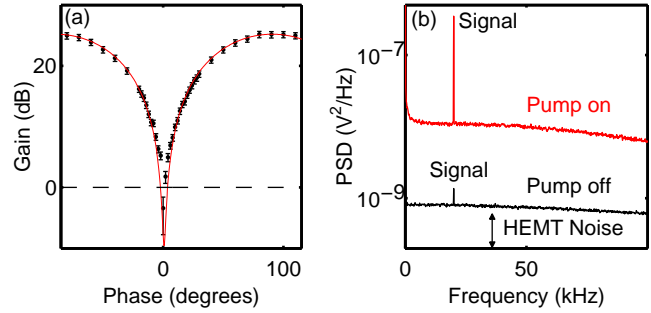}
\caption{\label{fig:Fig4}Performance of the amplifier in degenerate
mode. (a) Gain as a function of the phase between the pump and the
signal (points) and prediction (line) from Ref.
\onlinecite{yurke2006} ($P=0.9P_c$). (b) Power spectral density of
the demodulator output for the cases when the pump is on
($P=0.95P_c$) and off. The gain in this case is $630 \pm 30$ ($28
\pm0.2$~dB). The applied signal power is $1.6\pm 0.7
\times10^{-20}$~watts.}
\end{figure}

For $f_{res}=6.95$~GHz ($\Phi=0.2\Phi_0$), we have also operated the
JPA in a doubly degenerate mode where the pump and the signal
frequencies are the same. In this mode, the gain of the parametric
amplifier is sensitive to the phase between the pump and the signal.
To measure this phase dependence, we amplitude modulate the signal
at $20$~kHz and adjust the phase of the pump relative to the signal.
We define the gain as the ratio of the AM modulation sideband power
with the pump on and pump off. Because the local oscillator
frequency and the pump frequency are the same, the signal and
intermodulation tones are added at the output of the demodulator,
giving a total gain 3 dB larger than either DG or IG. In degenerate
mode, the gain can be 3 dB larger than in nondegenerate mode if the
phase between the pump and signal is tuned for maximum gain. The
phase dependence of the gain for a pump power close to $P_c$ is
shown in Fig.~\ref{fig:Fig4}(a); there it is evident that we see
deamplification, a hallmark of degenerate parametric amplifiers. In
Fig.~\ref{fig:Fig4}(b), we plot the power spectral density (PSD) of
the demodulated signal with the pump off and pump on for
$P=0.95P_c$, where the signal-pump phase has been adjusted for
maximum gain ($28 \pm0.2$~dB). At this gain, the HEMT amplifier's
input noise is overwhelmed by the noise at the output of the
parametric amplifier, effectively improving the signal-to-noise
ratio (S/N) by $16\pm0.4$~dB. A definitive measurement of the noise
added by our parametric amplifier will require a calibrated noise
source. We have not yet completed this measurement. However, by
measuring the S/N with the pump off, we find that the noise referred
to the input of the HEMT is $T_N=12\pm5$~K. From the S/N improvement
with the pump on, we estimate the total noise referred to the input
of the JPA as $300 \pm 130$~mK. This value suggests that the
parametric amplifier adds an amount of noise comparable to the
vacuum noise ($hf_{res}/2k_{B}=166$~mK), which must be present at
the input of the JPA.

To demonstrate the tunability of the JPA, we also test the
performance of the amplifier at lower frequencies. For example, for
$\Phi=0.35\Phi_0$, the resonance frequency is
$\omega_{0}/2\pi=5.203$ GHz. A similar analysis as the one described
for $\Phi=0.2\Phi_0$ gives the following parameters:
$\gamma_1/2\pi=0.95$~MHz, $\gamma_2/2\pi=0.85$~MHz, $P_c=0.5\pm
0.2$~fW, $\hbar K=-9 \pm 4\times 10^{-5}$ and
$\gamma_3=0.145\left|K\right|$. The increase in the nonlinear loss
degrades the performance of the amplifier, making the measured gains
smaller than the ones at $6.95$~GHz. The highest IG and DG observed
at this frequency are both $12$~dB.

Although the power-handling capacity of this device is low (critical
powers of the order of a few femtowatts), its performance is
appropriate for amplifying the signals generated by superconducting
qubits. By virtue of the tunability of our amplifier's band, it can
be brought into resonance with a second high-Q superconducting
resonator used to study superconducting qubits as in Refs.
\onlinecite{wallraff2004} and \onlinecite{schuster2007}. For more
general applications where larger signals need to be amplified,
similar parametric amplifiers could be used if the critical current
of the SQUIDs is made larger.

In conclusion, we have demonstrated a widely tunable parametric
amplifier based on a coplanar waveguide resonator whose inner
conductor is made from a SQUID array. We have observed tunability
over an octave and gains as high as $28$~dB. Although the resonator
is composed of discrete elements, its behaviour is well described by
a continuum theory of parametric amplification.\cite{yurke2006}
Finally we have demonstrated that the JPA is 16~dB more sensitive to
a weak microwave signal than a low-noise HEMT amplifier, suggesting
that the JPA adds less than twice the vacuum noise.

The authors thank S.~M. Girvin for valuable conversations. K.~W.
Lehnert is a member of NIST's Quantum Physics Division.

\bibliographystyle{apsrev}
%\bibliography{aplparaamp_ver14}

\end{document}